\documentclass[conference]{IEEEtran}
\IEEEoverridecommandlockouts
\usepackage{amsmath,amssymb,amsfonts}
\usepackage[utf8]{inputenc}
\usepackage{algorithm}
\usepackage[noend]{algpseudocode}
\algrenewcommand\algorithmicrequire{\textbf{Input:}}
\algrenewcommand\algorithmicensure{\textbf{Output:}}
\usepackage{array}
\usepackage{textcomp}
\usepackage{url}
\usepackage{graphics}
\usepackage{graphicx}
\usepackage{verbatim}
\usepackage{cite} %multi cite
\usepackage{siunitx}
\usepackage{booktabs, makecell}
\ifCLASSOPTIONcompsoc
    \usepackage[caption=false, font=normalsize, labelfont=sf, textfont=sf]{subfig}
\else
    \usepackage[caption=false, font=footnotesize]{subfig}
\fi
\usepackage{nomencl}
\usepackage{multirow,tabularx}
\usepackage{etoolbox}
\usepackage{array,multirow}
\usepackage{colortbl}
\usepackage{mathtools}
\usepackage{mdwtab}
\makeatletter
\newcommand*{\rom}[1]{\expandafter\@slowromancap\romannumeral #1@}
\makeatother
\usepackage{float}
\graphicspath{{./Figures/}}
\usepackage{comment}
\usepackage{gensymb} %for degrees symbol
\usepackage{bm} % bold and italic at once inside eqs
\usepackage{cleveref} %multi ref
\usepackage[table]{xcolor}
\makeatletter
\newcommand{\linebreakand}{%
  \end{@IEEEauthorhalign}
  \hfill\mbox{}\par
  \mbox{}\hfill\begin{@IEEEauthorhalign}
}
\makeatother

\begin{document}

\title{A Modified Sequence-to-point HVAC Load Disaggregation Algorithm\\
%{\footnotesize \textsuperscript}
}

\author{\IEEEauthorblockN{Kai Ye, Hyeonjin Kim, Yi Hu, Ning Lu}
\IEEEauthorblockA{
%\textit{Department of Electrical and Computer Engineering} \\
\textit{North Carolina State University}\\
Raleigh, NC 27606, USA \\
\{kye3, hkim66, yhu28, nlu2\}@ncsu.edu}
\and
\IEEEauthorblockN{Di Wu}
\IEEEauthorblockA{\textit{Pacific Northwest National Laboratory} \\
Richland, WA 99352, USA \\
di.wu@pnnl.gov}
\and
\IEEEauthorblockN{PJ Rehm}
\IEEEauthorblockA{\textit{ElectriCities of North Carolina Inc.} \\
Raleigh, NC 27604, USA \\
prehm@electricities.org}
}

\maketitle

\begin{abstract}
This paper presents a modified sequence-to-point (S2P) algorithm for disaggregating the heat, ventilation, and air conditioning (HVAC) load from the total building electricity consumption. The original S2P model is convolutional neural network (CNN) based, which uses load profiles as inputs. We propose three modifications. First, the input convolution layer is changed from 1D to 2D so that normalized temperature profiles are also used as inputs to the S2P model.
Second, a drop-out layer is added to improve adaptability and generalizability so that the model trained in one area can be transferred to other geographical areas without labelled HVAC data. 
Third, a fine-tuning process is proposed for areas with a small amount of labelled HVAC data so that the pre-trained S2P model can be fine-tuned to achieve higher disaggregation accuracy (i.e., better transferability) in other areas. 
The model is first trained and tested using smart meter and sub-metered HVAC data collected in Austin, Texas.  Then, the trained model is tested on two other areas: Boulder, Colorado and San Diego, California. Simulation results show that the proposed modified S2P algorithm outperforms the original S2P model and the support-vector machine based approach in accuracy, adaptability, and transferability.  
\end{abstract}


\section{Introduction}
Heat, ventilation, and air conditioning  (HVAC) loads are widely used demand response (DR) resources nowadays. This is because HVAC loads account for approximately 30\% of residential and 40\% of commercial building electricity consumption\cite{file2015res,file2015commercial}. In practice, only the total building electricity consumption is metered. Thus, load service providers and utility engineers mainly rely on HVAC load disaggregation algorithms for conducting DR potential studies.  

There are two categories of HVAC load disaggregation methods: \emph{unsupervised} and \emph{supervised}. Unsupervised methods do not require sub-metered data (i.e., labelled data). For example, in \cite{jacobs2020unsupervised}, a mixed-integer linear programming method is proposed for identifying appliance on/off state and consumption patterns using correlation and optimization. Supervised methods, such as support vector machine (SVM), decision tree, and k-Nearest Neighbors, require sub-metered data to extract features and train the model using those features\cite{angelis2022nilm}. In recent years, deep-learning methods (e.g., convolutional neural network (CNN) and long short-term memory (LSTM)) have been applied to HVAC disaggregation and achieved better accuracy\cite{cho2018non}. One advantage offered by CNN is that 
% manually extracting features is unnecessary because 
latent features of HVAC consumption patterns can be automatically captured by CNN layers. When abundant labelled data become available, supervised approaches normally outperform unsupervised approaches ones, and supervised machine-learning-based approaches are often used. 

In recent years, supervised machine learning methods have achieved excellent performance in non-intrusive load monitoring tasks \cite{schirmer2022non}. A major disadvantage is the requirement of sub-metered, high-resolution data as labelled inputs. In practice, only low-resolution (e.g., sampling rate at 15 minutes or lower) smart meter data are available. In addition, smart meters only record the total building electricity consumption and sub-metered HVAC data is usually not available for training HVAC load disaggregation algorithms.

Another disadvantage is that the generalizability of the  model is usually poor. Because the training is site-specific, disaggregation accuracy can drop significantly when applied at other locations or to different types of customers. In \cite{d2019transfer}, Michele D’Incecco \emph{et al}. introduced transfer learning to generalize the Sequence-to-point (S2P) CNN-based model to different appliances and locations by fine-tuning. However, model training and fine-tuning still require high-resolution, sub-metered data at new sites. 

To overcome these disadvantages, we propose a modified S2P algorithm to use low-resolution smart meter data as inputs for HVAC load disaggregation. Our contributions are threefold. \emph{First}, we modify the convolution layer from 1D to 2D to use both temperature and load as inputs to the S2P model.
\emph{Second}, we add a drop-out layer to improve model adaptability and generalizability so that models trained in one area can be transferred to other areas without labelled data. 
\emph{Third}, a fine-tuning process is proposed for areas with small amounts of labelled data so that the pre-trained S2P model can be fine-tuned to achieve higher disaggregation accuracy and thereby better transferability. 

The proposed model is first trained and tested using smart meter and sub-metered HVAC data collected in Austin, Texas.  Then, the pre-trained model is tested in two other areas: Boulder, Colorado and San Diego, California. Simulation results show that the proposed modified S2P algorithm outperforms the original S2P model and the support-vector machine based approach in terms of accuracy, adaptability, and transferability.

\section{Methodology}\label{sec:method}
In this section, we present the S2P method, hyper parameter selection, the transfer learning mechanism, and performance evaluation metrics.
%\subsection{Convolutional Neural Network Model}
\subsection{An Overview of the Modified S2P Algorithm}\label{sec:overview}
\begin{figure}[htbp]
%\centerline{\includegraphics[width=\textwidth, height=2in]{model.png}}
\centerline{\includegraphics[width=3.5in]{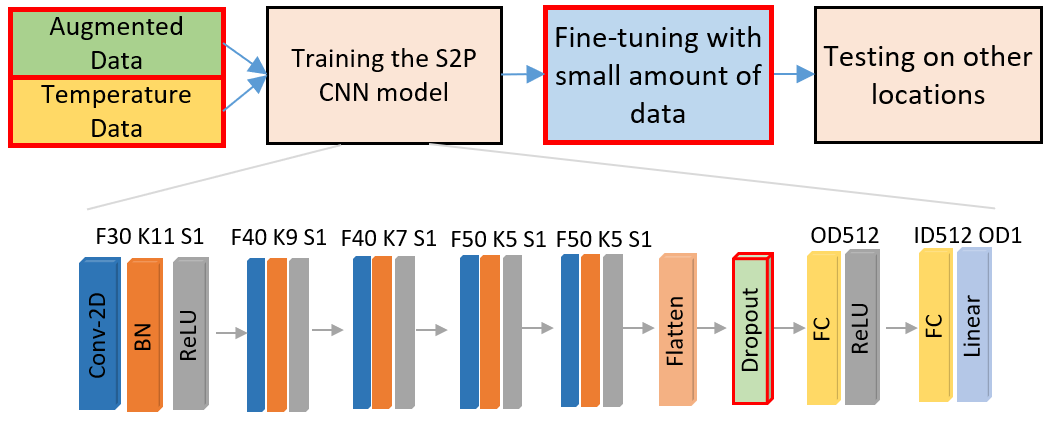}}
\vspace{-.5cm}
\caption{Architecture of the modified S2P HVAC load disaggregation algorithm. Corresponding input dimension (ID), output dimension (OD), filter(F), kernel size (K), stride (S), padding (P), output padding (OP) for each layer. “Conv-2D” refers to 2D CNN layer. “BN” refers to batch normalization layer, “FC” refers to fully connected layer, 'ReLU' refers to rectified linear unit.}
\label{fig1}
\vspace{-.3cm}
\end{figure}

\begin{figure}[htbp]
%\centerline{\includegraphics[width=\textwidth, height=2in]{model.png}}
\centerline{\includegraphics[width=3in]{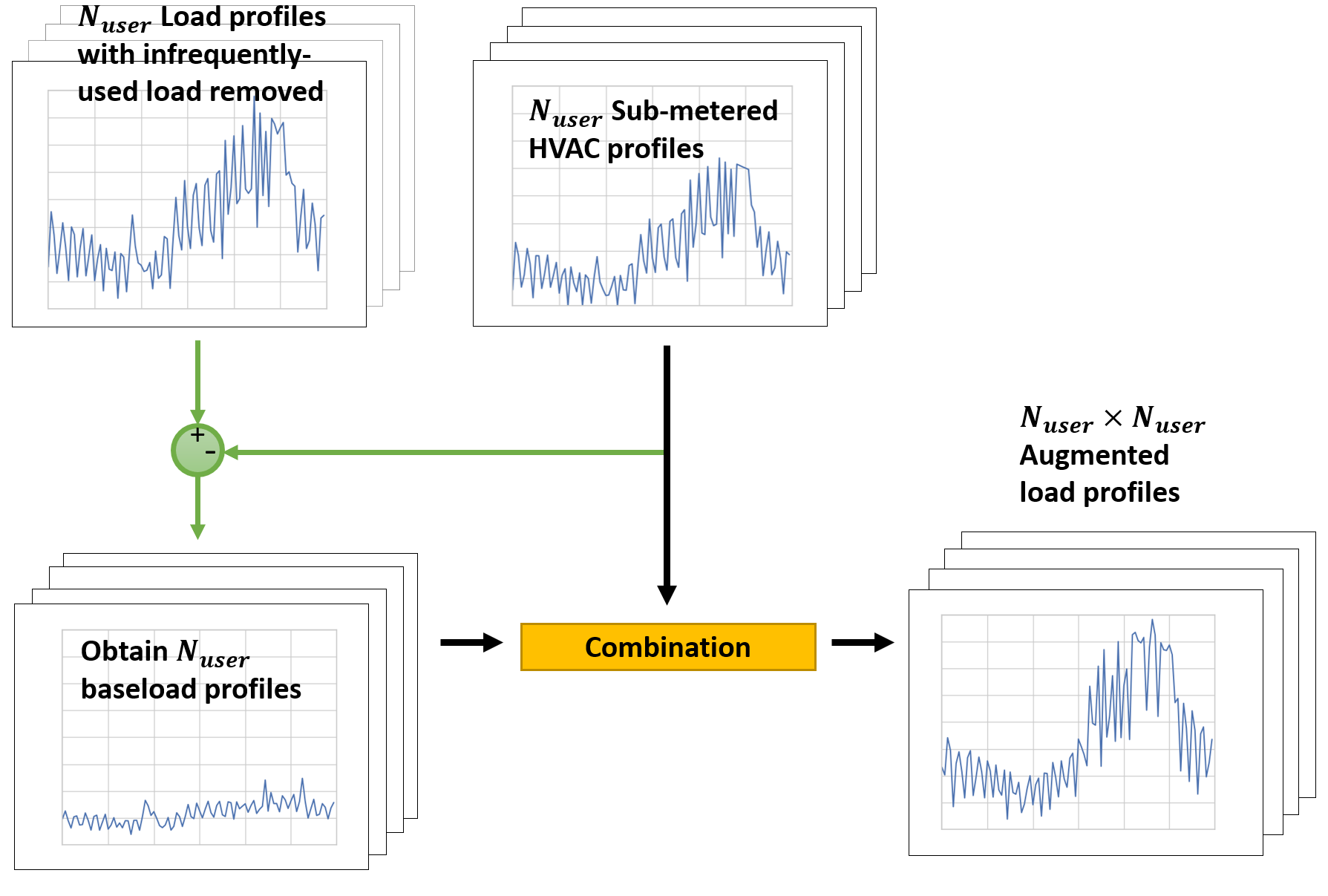}}
\vspace{-.3cm}
\caption{Data augmentation workflow.}
\label{fig2}
\vspace{-.5cm}
\end{figure}

As shown in Fig. \ref{fig1}, the modified S2P model structure includes three main processes: data augmentation, training and testing the model on one location, transfer learning (port the pre-trained model to other locations) with fine-tuning. The data augmentation process is illustrated in Fig. \ref{fig2}. First, the infrequently-used loads (i.e., water heater and dryer loads) are removed from the total load profile
using the algorithm introduced in \cite{liang2019hvac}. 
Then, the sub-metered HVAC data is removed from the residual load profile to obtain the base-load profiles. Next, the augmented load profiles are generated by shuffling the HVAC profiles against each base load profile. After data augmentation, the original $N_{\mathrm{user}}$ yearly 1-minute load profiles are expanded to $N_{\mathrm{user}}\times N_{\mathrm{user}}$ yearly 1-minute load profiles. Then, the augmented load profiles and their corresponding ambient temperature profiles are used as the labeled data for training the modified S2P model. The trained S2P model is first tested for the same location. Then, it is ported to two other locations, where we compare the pre-trained model performance for two cases: with and without fine-tuning. Because the original S2P model structure and the problem formulation is introduced in \cite{zhang2018sequence}, in this paper, we only describe the modified components (see the red boxes in Fig. \ref{fig1}). 

\subsection{Sequence-to-point CNN model}\label{sec:cnn}
Assume that there are $N_{\mathrm{user}}$ users in the training set. Select 90 days in the three summer months from each user. Assume the data sampling interval is $\Delta t$ minutes, then, there are $N_{\mathrm{data}} = 90\times24\times60/\Delta t$ training samples in the training set of user $i$.

\begin{figure}[htbp]
\vspace{-.3cm}
%\centerline{\includegraphics[width=\textwidth, height=2in]{model.png}}
\centerline{\includegraphics[width=3.5in]{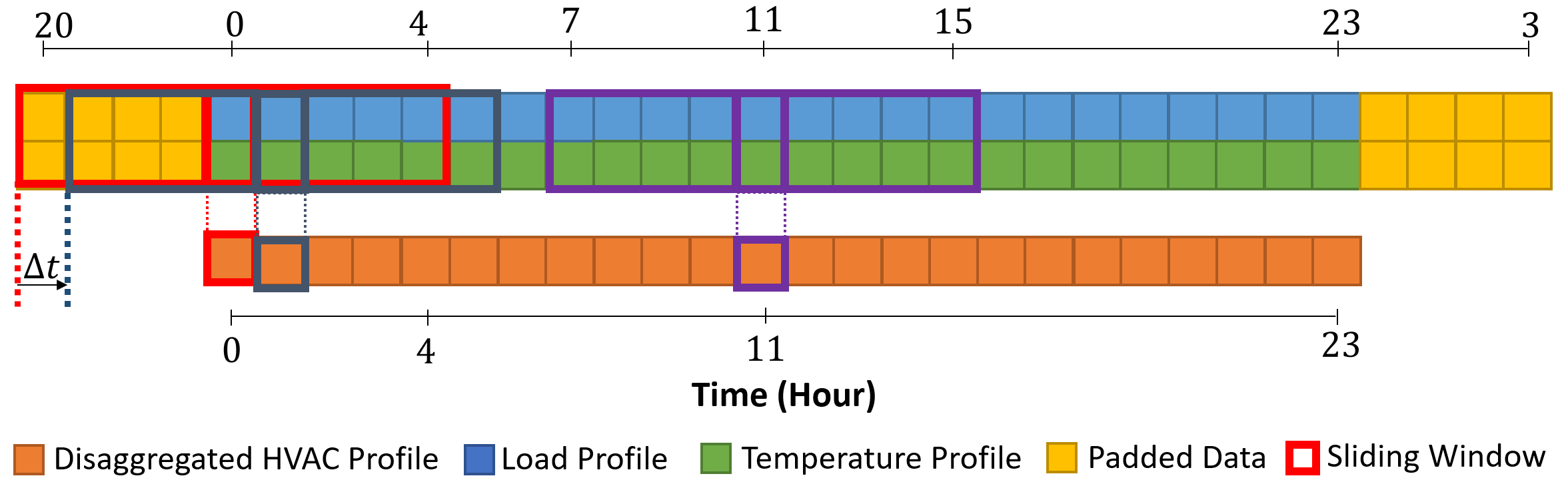}}
\vspace{-.3cm}
\caption{Illustration of the model input and output.}
\label{fig3}
\vspace{-.2cm}
\end{figure}

The S2P model (See Fig. \ref{fig1}) disaggregates the HVAC load one data point at a time. As shown in Fig. \ref{fig3}, 
%at each time step $t$, the model takes load and temperature time sequence (data points in the red box), $[t-K, t+]$, as inputs and make one prediction of the HVAC power consumption at $t$. 
to identify the actual HVAC load at time $t$, $P^{\mathrm{HVAC}}_{i,t}$ for the $i^{\mathrm{th}}$ user ($i\in[1,...,N_{\mathrm{user}}]$), the S2P model selects $K$ data points before and after time $t$ ($t\in[1,...,N_{\mathrm{data}}$]) from the $i^{\mathrm{th}}$ user ($P_i$) and the corresponding temperature profile ($T_i$), respectively. So the inputs of the model are
\begin{equation}
    {P}_{i,t}^{\rm{Input}} = P_i(t-K:t+K)    \\
\end{equation}
\begin{equation}
    {T}_{i,t}^{\rm{Input}} = T_i(t-K:t+K)\\
\end{equation}
The output is $\hat{P}^{\mathrm{HVAC}}_{i,t}$. Thus, assume that the data sampling rate is $\Delta t=1$ hour and $t=11$. If $K=4$, then 4-hour ahead and 4-hour after 11:00 will be selected as inputs to predict the HVAC consumption at 11:00. Thus, to disaggregate the HVAC load from hour 1 to hour 24, we need to pad 4-hour data at the beginning and at the end of the day (yellow boxes in Fig. \ref{fig3}). By sliding the input window from hour 1 to hour 24 one data point at a time, the HVAC load can then be generated for each time interval in a day. 

Note that the original S2P model only uses the total household load consumption as inputs. By introducing the temperature as inputs, the modified S2P model can identify the HVAC load temperature dependency to further improve the model generalizability and identification accuracy.

The obtained HVAC load profile is post-processed by 
\begin{equation} \label{eq_bound}
%\vspace{-0.5cm}
    \Tilde{P}^{\mathrm{HVAC}}_{i,t}= 
\begin{cases}
    P^{\mathrm{Input}}_{i,t},& \text{if } \hat{P}^{\mathrm{HVAC}}_{i,t}\geq P^{\mathrm{Input}}_{i,t}\\
    \hat{P}^{\mathrm{HVAC}}_{i,t},& \text{if } P^{\mathrm{Input}}_{i,t} \geq \hat{P}^{\mathrm{HVAC}}_{i,t}\geq \epsilon\\
    0,              & \text{otherwise}
\end{cases} 
%\vspace{-0.3cm}
\end{equation}
where $\epsilon$ is the cutoff power threshold for setting very small HVAC consumption to be zero. This process can remove very small values and reinforce that the HVAC unit consumes less than the total power.

\subsection{Model Generalization and Transfer Learning} \label{sec:finetuning}

To use the pre-trained model in other areas without sub-metered HVAC data, we made two modifications to the original S2P model structure: including the temperature profile as inputs and adding a dropout layer. Assume that HVAC power consumption is highly correlated with ambient temperature variations and customer preferences of thermostat settings are similar across the United States. Then, by adding localized ambient temperature as inputs, we hope to improve the generalizability of the model. Adding a dropout layer prevents over-fitting and subsequently provides the model with excellent generalization ability.

\begin{figure}[htbp]
\vspace{-0.4cm}
\centerline{\includegraphics[width=3.5in]{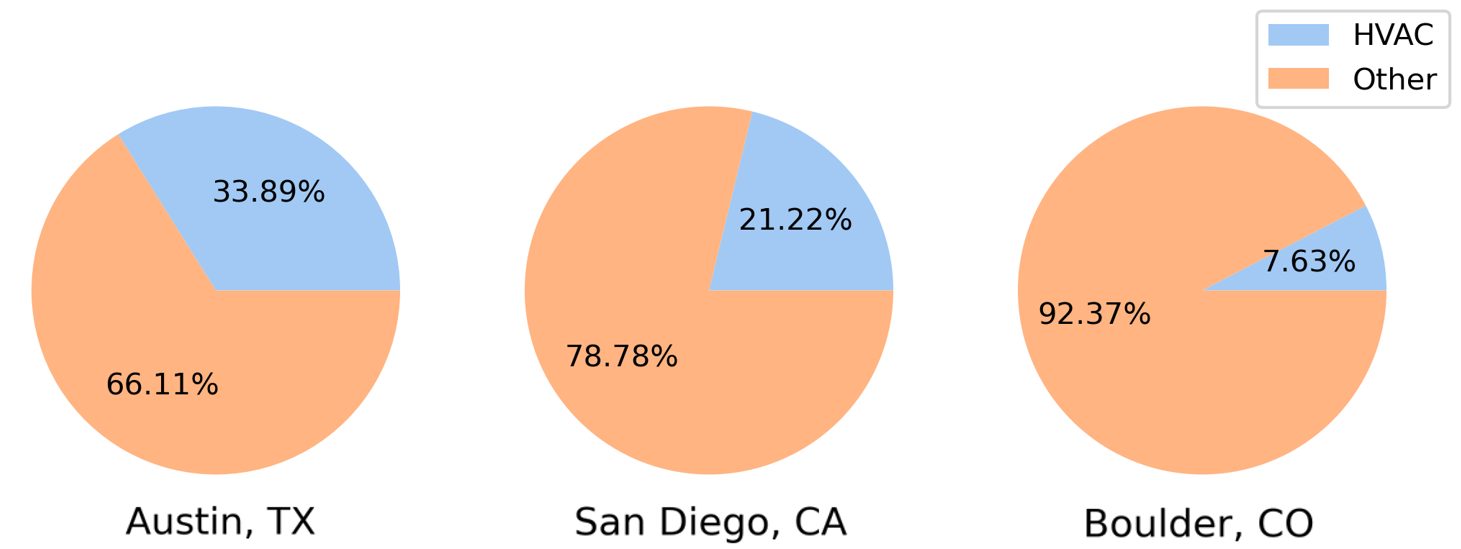}}
\vspace{-0.2cm}
\caption{Average household HVAC energy consumption ratios (Summer 2015).}
\label{energy}
\vspace{-0.2cm}
\end{figure}

To verify the above hypothesis, we test the pre-trained model on sub-metered data sets collected from two other sites: Boulder, Colorado and San Diego, California for verification. However, as shown in Fig. \ref{energy}, the HVAC electricity consumption ratio varies in different areas and the average HVAC energy consumption ratio in Austin is significantly higher than that of San Diego or Boulder. This means that the load composition may be significantly different in different areas, which can affect the generalization. 

\begin{figure}[htbp]
\vspace{-0.3cm}
\centerline{\includegraphics[width=3.5in]{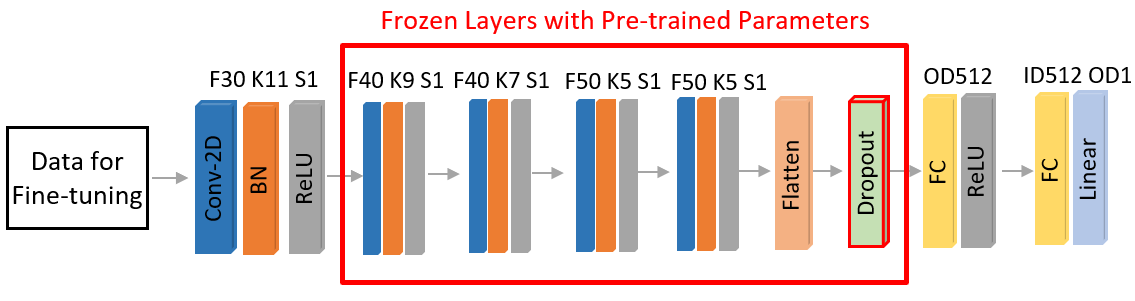}}
\vspace{-0.5cm}
\caption{Fine-tuning the pre-trained modified S2P model.}
\label{finetune}
\vspace{-0.3cm}
\end{figure}

To mitigate the impact of such differences in end use load consumption compositions, we apply the transfer learning to fine-tune the pre-trained model. As introduced in\cite{d2019transfer}, transfer learning requires using a small amount of labelled data collected at the new sites as inputs. The pre-trained model is fine-tuned by training only the first CNN layer and the last fully connected layer while keeping parameters of all remaining layers (included in the red box) fixed (see Fig. \ref{finetune}).

\subsection{Loss Function and Hyper Parameter Selection}\label{sec:hyper_para}
The loss function is calculated as
\begin{equation}
    {L}_p = \frac{1}{N_{\mathrm{data}}} \sum_{t=1}^{N_{\mathrm{data}}}\| \hat{P}^{\mathrm{HVAC}}_{t} -P^{\mathrm{HVAC}}_{t}\| ^2_2    \\
\end{equation}

The parameters for each S2P layer are shown in Fig. \ref{fig1}. The model is implemented in Tensorflow using the ADAM optimizer. The hyper parameters are listed in Table \ref{tabhyper}.

\begin{table}[htbp]
\vspace{-0.2cm}
\caption{Hyper Parameters of the Modified S2P Model}
\vspace{-0.5cm}
\begin{center}
\begin{tabular}{c|c}
\hline\hline
\textbf{Input window size}&\textbf{Table} \\
\hline
Number of epochs & 30  \\
\hline
 Input window length $(2K+1)$& 33 \\
\hline
Batch size & 1000  \\
\hline
Learning rate & 0.005 \\
\hline
Dropout rate & 0.4 \\
\hline
Transfer learning rate & 0.001 \\
\hline
Transfer learning epoches & 15 \\
\hline\hline
\end{tabular}
\label{tabhyper}
\end{center}
\vspace{-0.5cm}
\end{table}

\subsection{Performance Evaluation Metrics}
To analyze the accuracy of the disaggregated load profile, normalized mean absolute error ($n$MAE) can be computed by 

\begin{equation}
n\textbf{MAE}=\frac{1}{N} \cdot \sum\limits_{t = 1}^{N} {\frac{{\left| {\Tilde{P}^{\mathrm{HVAC}}_{t} - {P}^{\mathrm{HVAC}}_{t} } \right|}}{{P_t^{\mathrm{rated}}}}} \label{eq1}    
\end{equation}
where $N$ is the total number of data points in 90 days.

Besides the point-to-point match of the disaggregated HVAC load curves, the hourly energy error is also an important metrics used by utilities to estimate the HVAC potential for providing DR. Thus, hourly normalized energy error ($n$EE) is calculated as
\begin{equation}
n\textbf{EE} = {\frac {\left| \sum\limits_{t = 1}^{N_{\mathrm{hour}}} {\Tilde{P}^{\mathrm{HVAC}}_{t}} - \sum\limits_{t = 1}^{N_{\mathrm{hour}}}{P}^{\mathrm{HVAC}}_{t} \right|} {{\sum\limits_{t = 1}^{N_{\mathrm{hour}}} {{P}^{\mathrm{rated}}_{t}}}}}  \label{eq2}
\end{equation}
where $N_{\mathrm{hour}}$ is the total number of data points in an hour. 
$n\textbf{EE}$ is calculated for every hour in 90-day summer days for each household so  HVAC hourly energy error distributions from hour 1 to 24 can be calculated.

\section{Simulation Results}
This section presents simulation results and performance comparisons.

\subsection{Data Preparation}
The data sets used in this study were collected by Pecan Street Inc. from 1070 residential users in New York, Colorado, California, and Texas \cite{pecan}. The data sets include 1-minute electricity consumption of the total household and appliances (e.g., HVAC, water heater, and dryer). The 1-minute data is down-sampled to 15-minute to match the typical smart meter data resolution. In this paper, due to space limit, we select users with only one HVAC unit and illustrate the algorithm performance on cooling load disaggregation so that 230 users and 90 days in Summer 2015 are selected. Thus, $N_{\mathrm{data}}=24\times4$, $N_{\mathrm{hour}}=90\times24$, and  $N=90\times24\times4$.

As a first step, all 230 sets of load profiles are normalized by using the maximum power consumption of the whole set as the base, where $P_{\mathrm{base}}=27.24$ kW. Thus, all load profiles are converted into per unit values (between 0 and 1). Similarly, the temperature profiles are also normalized so the values are between 0 and 1, making the load and temperature data sets in the same value range. 

Next, the modified S2P model is trained on the augmented data set using hyper parameters specified in Section \ref{sec:hyper_para}. Note that data augmentation is conducted following the method introduced in Section \ref{sec:overview}, which uses the original data sets from 150 users in Austin as inputs.  The trained model is first tested on the remaining 50 users in the same location (Austin, TX). Then, the pre-trained model is tested on data sets collected in two other locutions: 20 users in Boulder, CO and 10 users in San Diego, CA.

\subsection{Model architecture Selection}
This section compares the model performance for cases with different numbers of convolutional layers and with/without the dropout layer. As shown in Table \ref{tabmodel}, using more CNN layers can potentially increase the performance of the model by extracting hidden features. However, it can also lead to over-fitting, especially when the training dataset is small. Adding a dropout layer can help alleviate overfitting and achieve better generalization in unseen data sets.   
Using 5 convolutional layers with dropout achieves performance improvements in all three locations, showing a good adaptability and generalizability when the pre-trained model is transferred to another area. The results show that by adding the dropout layer to the original S2P structure, the model performance has been significantly improved.

\begin{table}[htbp]
\vspace{-0.45cm}
\caption{Performance Comparison for Different Model Structures}
\vspace{-0.5cm}
\begin{center}
\begin{tabular}
% {0.6\textwidth}{XXXXXXXX}
{ccccccc}
% {m{3.5em}m{1em}m{3em}m{3em}m{3em}m{3em}m{3em}m{3em}}
\toprule 
\toprule
% \multicolumn{2}{c}{}&\textbf{4-layer}&\textbf{4-layer w/D} &\textbf{5-layer} &\textbf{5-layer w/D}&\textbf{6-layer} &\textbf{6-layer w/D}\\
 &\multicolumn{3}{c}{\textbf{$n$MAE(\%)}} & \multicolumn{3}{c}{\textbf{$n$EE(\%)}}\\
 \cmidrule{2-4}  \cmidrule{5-7}
 & \textbf{TX} & \textbf{CO} & \textbf{CA}&  \textbf{TX} & \textbf{CO} & \textbf{CA}\\
\hline
\\[-1em]
4-layer & 9.11 &10.44 & 5.48& 4.21 &4.95 & 2.43\\

\\[-1em]
4-layer with dropout& 8.58 &10.13 & 5.29& 3.97 &4.35 & 2.24\\

\\[-1em]
5-layer & 9.44 &10.61 & 5.83& 4.34 &4.87& 2.46\\

\\[-1em]
\textbf{5-layer with dropout} & \textbf{7.17} &\textbf{9.28} & \textbf{4.40}& \textbf{3.51} &\textbf{4.10} & \textbf{1.89}\\

\\[-1em]
6-layer & 8.77 &10.47 &5.43& 3.82 &5.17& 2.20\\

\\[-1em]
6-layer with dropout & 8.82 &10.11 & 5.33& 3.64 &4.42 & 2.18\\
% &CO & 1 &1 &1& 1 &1 & 1\\
% &CA& 1 &1 &1& 1 &1 & 1\\
% \hline
% \multirow{3}{4em}{\textbf{$n$EE(\%)}} &TX  & 1 &1 & 1& 1 &1 & 1\\
% &CO& 1 &1 &1& 1 &1 & 1\\
% &CA& 1 &1 &1& 1 &1 & 1\\
% \hline
\bottomrule
\bottomrule % <-- Bottomrule here
\end{tabular}
\label{tabmodel}
\end{center}
\vspace{-0.55cm}
\end{table}

\subsection{Input Window Selection at Different Data Granularity}
\begin{figure}[htbp]
\vspace{-0.4cm}
\centerline{\includegraphics[width=3.2in]{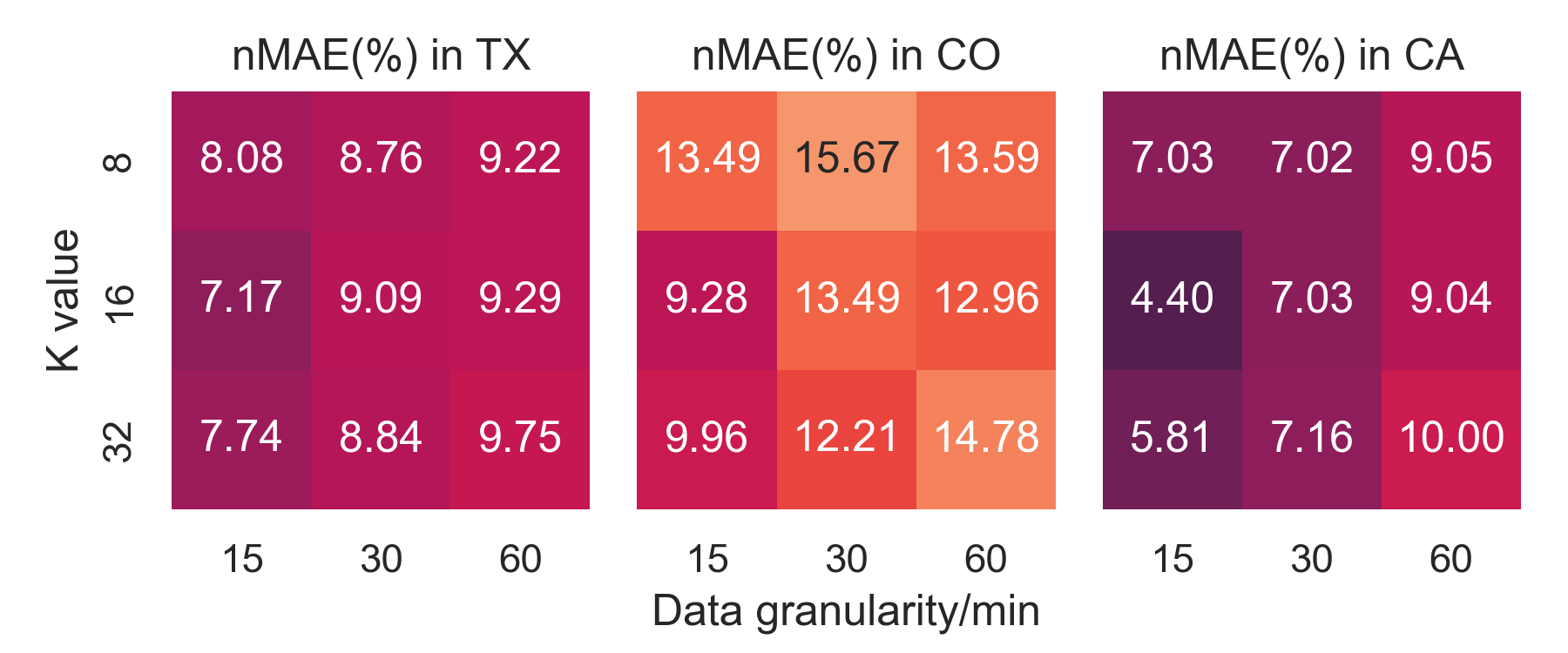}}
\vspace{-0.5cm}
\caption{Performance comparison when selecting different input data resolutions and window lengths.}
\label{mdgra}
\vspace{-0.45cm}
\end{figure}

As in some area, the smart meter data resolution can be 30-minute or 60-minute instead of 15-minute. Thus, we test the model performance for different input data granularity and for different input data window length ($2K+1$). As shown in Fig. \ref{mdgra}, the down-sampling of data inevitably resulted in information loss. The accuracy degradation is small in Austin, TX because there are abundant labelled training data.  However, the transferability decreased quickly when data resolution is lower. This is because if the data sampling interval is longer than 15-minute, it becomes harder to capture the HVAC on/off cycles, which are usually less than 30-minutes. 
Meanwhile, an appropriate choice of input window length is 4-hour before and 4-hour after time $t$ across different areas. The results also show that the HVAC on/off status is heavily correlated with ambient temperature variations instead of user pattern shifts because the user pattern shifts (e.g., weekdays and weekends) do not explicitly affect the model performance. 

\subsection{Efficacy of data augmentation and temperature encoding}
In this subsection, we compared  the proposed model with two benchmark models using data collected in Austin, Texas: SVM and the original S2P-CNN. We set up three cases to compare the performance improvements brought by data augmentation and adding temperature to the input: 1) using data sets collected from the 150 users for training, 2) using augmented data ($150\times150$ users) for training, and 3) using augmented data for training and using both load and temperature data as model inputs. The trained model is tested on the remaining 50 Austin users. 

\begin{figure}[htbp]
\vspace{-0.35cm}
\centering
\subfloat[]{\includegraphics[width=3.5in]{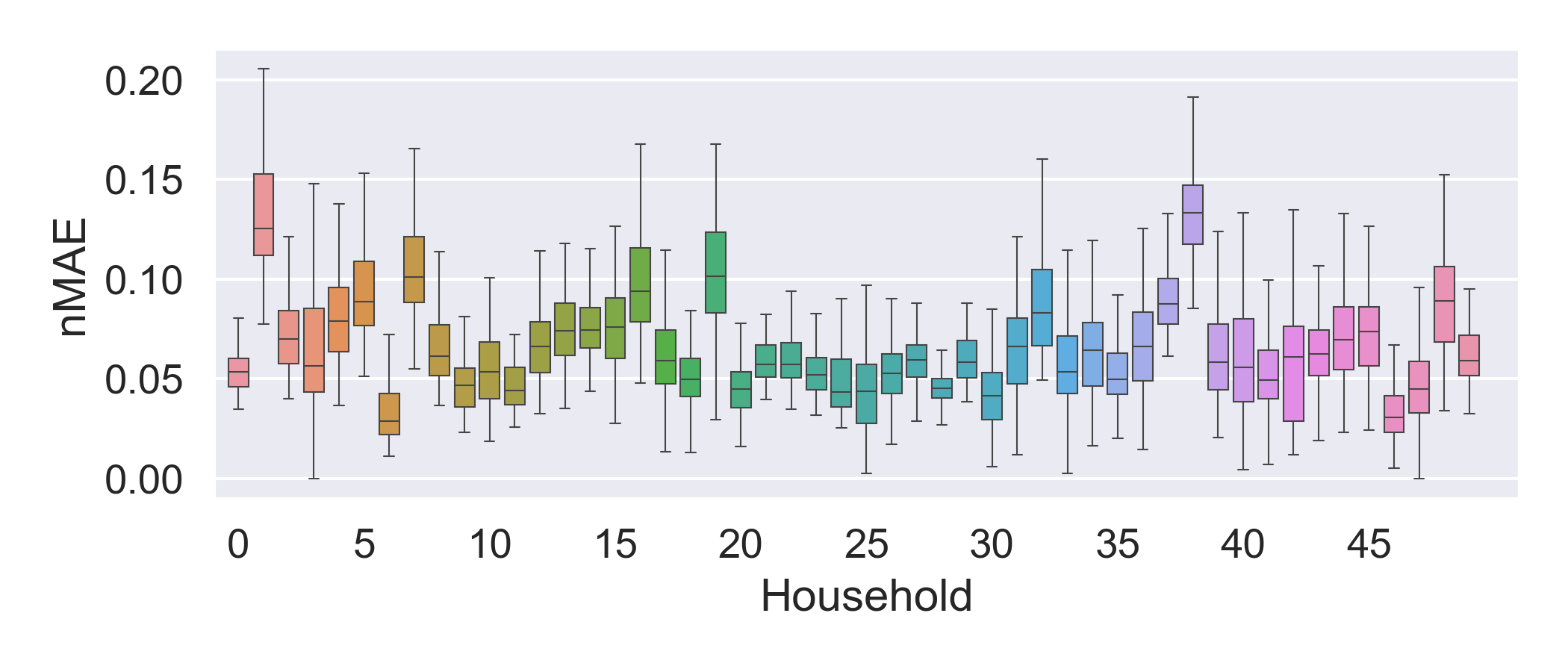}\label{fig7_1}}\\
\vspace{-0.3cm}
\subfloat[]{\includegraphics[width=3.5in]{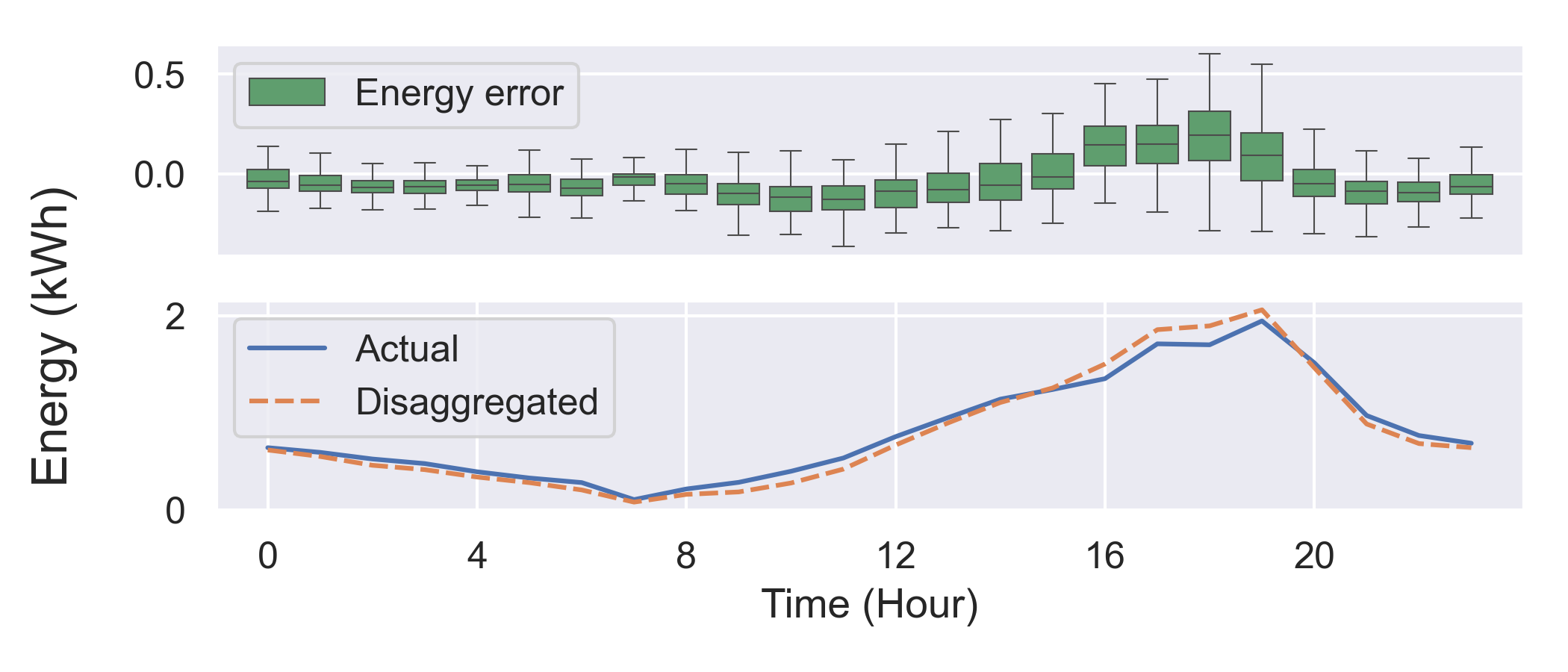}\label{fig7_2}}\\
\caption{(a) Daily $n$MAE distributions for 50 households in Texas. (b) Hourly average HVAC consumption (actual versus forecasted) and hourly energy error distributions for a typical household.}
\label{fig7}
\end{figure}

As shown in Table \ref{tabast}, data augmentation significantly reduces error variances (i.e., improved model consistency) and adding temperature to the inputs reduces the errors (i.e., improved accuracy). As shown in Fig. \ref{fig7}(a), the trained model has shown good performances (i.e, the average daily $n$MAE less than 10\%) for 46 out of 50 users. In Fig. \ref{fig7}(b), the lower plot shows the average disaggregated HVAC consumption compared to the actual value for a household with an average $n$MAE of 6.67\%. And the corresponding energy error (in kWh) distributions at each hour of the day is shown in the the upper plot of Fig. \ref{fig7}(b). Note that higher hourly energy errors occur between 16:00 and 19:00. This is because there are significant amounts of thermostatically-controlled cyclic loads in presence during those hours (e.g., refrigerators and some cooking loads). 

\begin{table}[htbp]
\vspace{-0.5cm}
		\caption{Performance Comparison \\(Point-to-point Errors and Hourly Energy Errors)}
\vspace{-0.5cm}
	\begin{center}
		\label{tabast}
		\begin{tabular}{ccccccc}
		\toprule
			\toprule % <-- Toprule here
&\multicolumn{2}{c}{\textbf{Benchmark Models}} & &\multicolumn{3}{c}{\textbf{Proposed Model}}\\
 \cmidrule{2-3}  \cmidrule{5-7}
& \textbf{SVM} & \textbf{S2P-CNN} & & \textbf{Case 1} & \textbf{Case 2} & \textbf{Case 3}\\
\midrule
$n$MAE (\%) & 13.09 &9.54 & & 8.54 & 8.44 & {\textbf{7.17}}\\
$n$EE (\%) & 11.36 & 6.47 & &4.37 & 4.54 & {\textbf{3.51}}\\
std($n$MAE) & 8.47 &4.25 & & 3.89 & 2.28 & {\textbf{2.85}}\\
std($n$EE) & 7.42 & 3.72 & & 2.39 & 2.47 & {\textbf{1.86}}\\
		    \bottomrule
			\bottomrule % <-- Bottomrule here
		\end{tabular}
	\end{center}
\vspace{-0.5cm}
\end{table}

\subsection{Effect of the transfer-learning based fine-tuning}
For new sites with limited amounts of sub-metered data, we can fine-tune the model for each household using the transfer learning algorithm introduced in Section \ref{sec:finetuning}. In this paper, we use one weeks' labelled data to demonstrate the efficacy of fine-tuning.

As shown in Table \ref{trans1}, the pre-trained model has shown satisfactory accuracy when directly used on new sites. The accuracy achieved at San Diego is close to that of Austin, while the accuracy in Boulder is lower. This may be caused by greater differences in weather patterns. Table \ref{trans1} results also show that fine-tuning can considerably reduce the disaggregation errors and error variances.

\begin{figure}[htbp]
\vspace{-0.4cm}
\centerline{\includegraphics[width=3.6in]{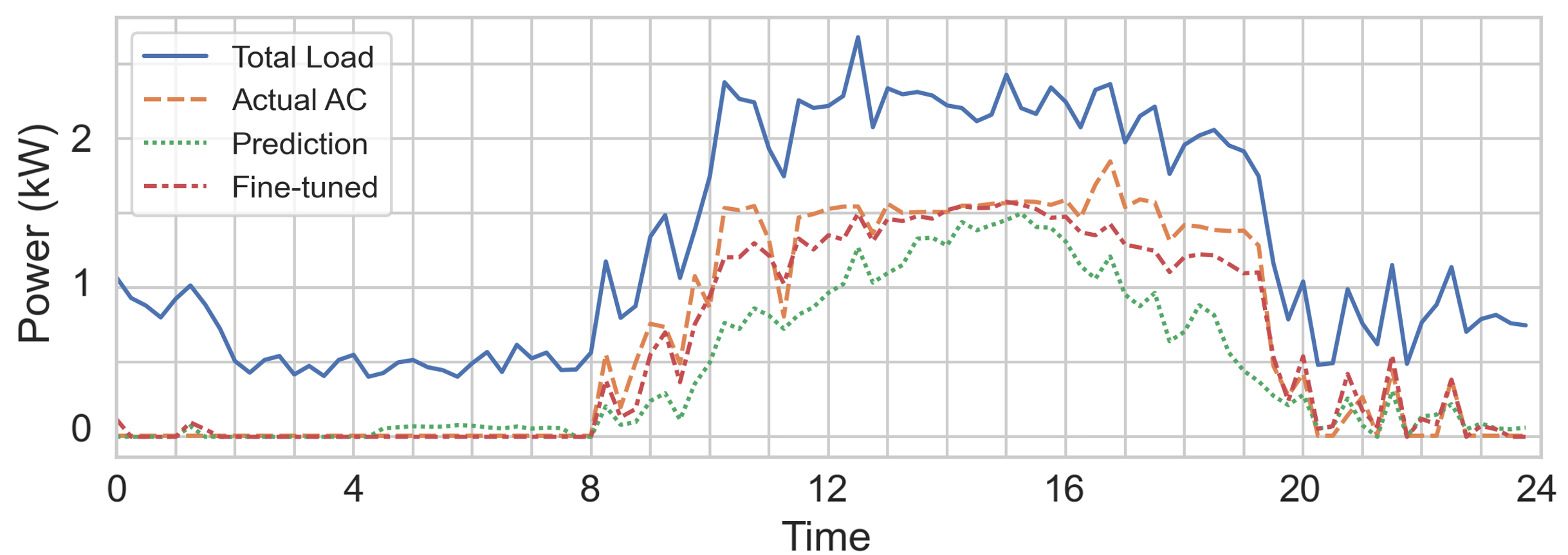}}
\vspace{-0.3cm}
\caption{An example of load disaggregation results for one user located at Boulder, CO.}
\label{maeuser}
\vspace{-0.2cm}
\end{figure}

\begin{figure}[htbp]
%\vspace{-0.5cm}
\centering
\subfloat[]{
    \includegraphics[width=3.2in]{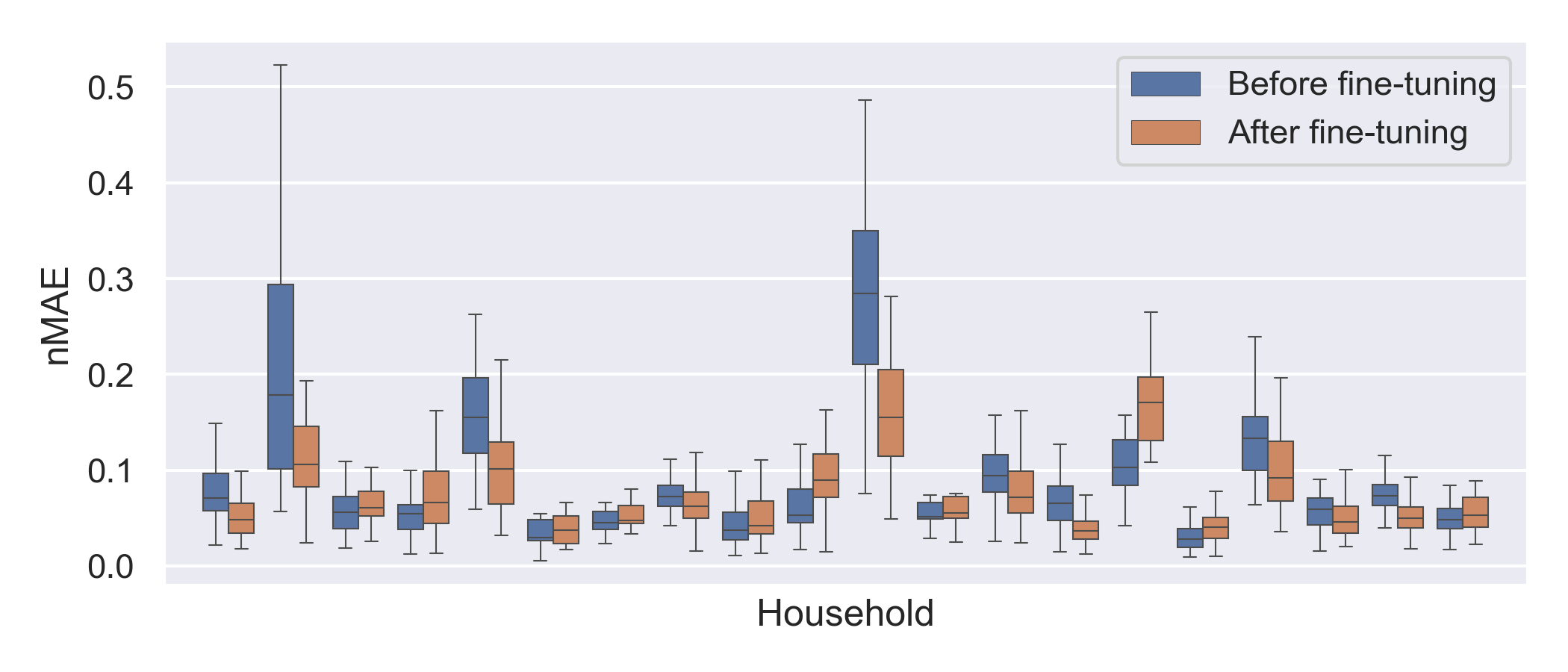}
    % \vspace{-0.2cm}
    \label{fig9_1}}\\
\vspace{-0.3cm}
\subfloat[]{\includegraphics[width=3.2in]{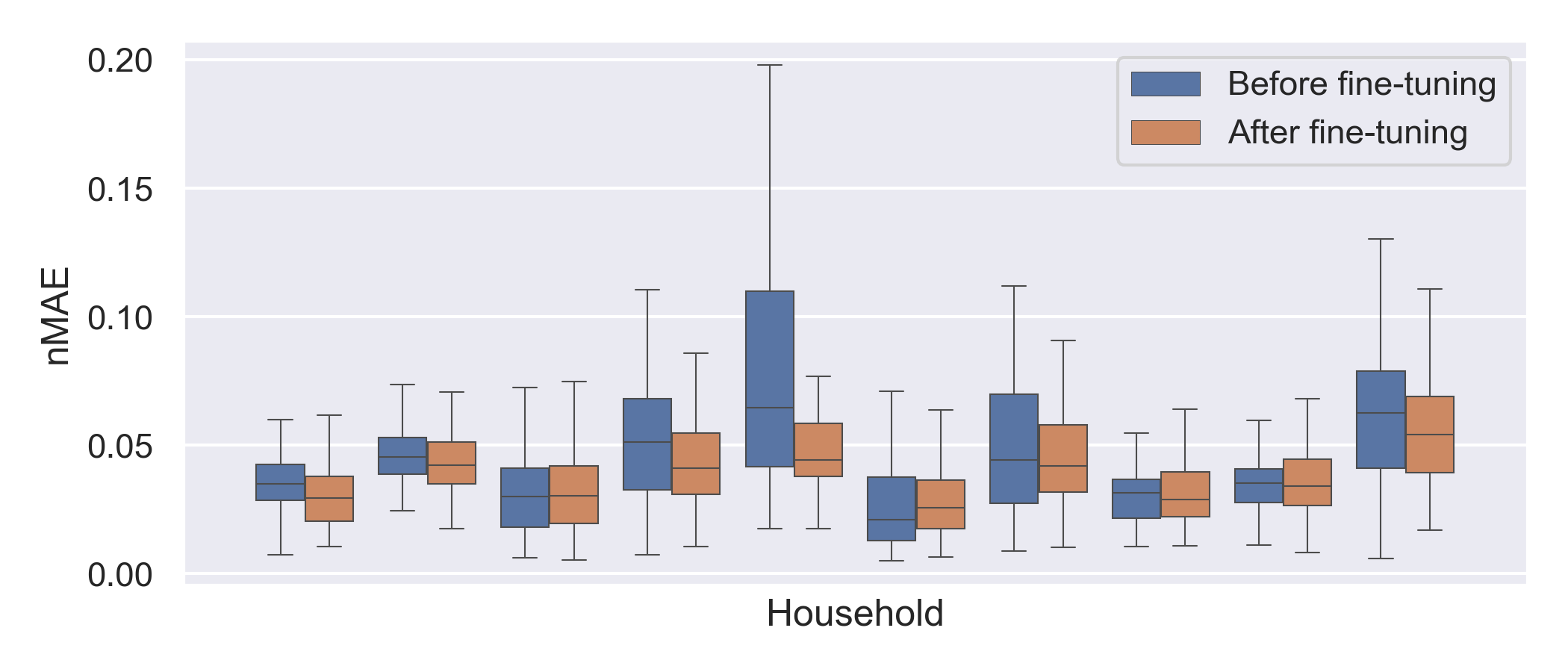}\label{fig9_2}}\\
\caption{Daily average normalized point-to-point error ($n$MAE) distributions (a) for 20 users in Boulder, CO. and (b) for 10 users in San Diego, CA.}
\label{fig9}
\vspace{-0.6cm}
\end{figure}

As shown in Fig. \ref{maeuser}, fine-tuning achieves noticeable improvements in HVAC disaggregation accuracy because it can identify some distinct user characteristics (e.g., the HVAC rated power). The point-to-point error distributions of each household before and after fine-tuning are shown in Fig. \ref{fig9}. The results show that the accuracy of the model can be significantly improved with fine-tuning for most houses, especially for the households where the pre-trained model without using fine-tuning performs poorly (e.g., nMAE greater than 10\%). 
For users where the pre-trained model achieves very good performance (i.e., nMAE less than 5\%), fine-tuning brings little or no improvement.  This implies that the key features of those users have already been successfully captured by the pre-trained model.

\begin{table}[htbp]
\vspace{-0.2cm}
	\begin{center}
		\caption{Performance Comparison for different locations \\ (20 users in Boulder, CO, and 10 users in California, CA)}
		\vspace{-0.2cm}
		\label{trans1}
		\begin{tabular}{ccccccc}
		\toprule
			\toprule % <-- Toprule here
\multirow{2}{*}{\textbf{Area}}& \multirow{2}{*}{\textbf{Metrics}}&\multicolumn{2}{c}{\textbf{Benchmark Models}} & &\multicolumn{2}{c}{\textbf{Modified S2P}}\\
 \cmidrule{3-4}  \cmidrule{6-7}
& & \textbf{SVM} & \textbf{S2P-CNN} & & \textbf{NoFT} & \textbf{WithFT} \\
\midrule
\multirow{2}{*}{CO} &$n$MAE (\%) & 23.78 &11.26 & & 9.28 & 7.90\\
& std($n$MAE) & 17.12 &7.61 & & 6.54& 3.73\\
\midrule
\multirow{2}{*}{CA} &$n$MAE (\%) & 19.82 &6.71 & & 4.40 & 4.52\\
& std($n$MAE) & 9.62 &3.55 & & 1.82 & 0.86\\
		    \bottomrule
			\bottomrule % <-- Bottomrule here
		\end{tabular}
	\end{center}
\vspace{-0.5cm}
\end{table}

\section{Conclusion and Future Work}
In this paper, we proposed a modified S2P CNN-based HVAC load disaggregation algorithm. We first demonstrated that by adding a dropout layer and using temperature data as inputs, the modified model shows significant improvement in disaggregation accuracy and consistency compared to previous methods in an area with labelled data sets. Then, we show that the model generalizability is also improved because the pre-trained model achieves satisfactory performance in other locations where there is no sub-metered HVAC load data for training the model. We further analyze the impact of input data resolution and show that both the disaggregation accuracy and the model transferability degrades when the data sampling interval is longer than 15-minute. To further improve the transferability, we show that using a small amount of labelled data at the new location to fine-tune the pre-trained model can significantly improve the model performance. Our future work is to combine a few complementary HVAC disaggregation methods together to further improving identification accuracy, results consistency, and generalizability.

\bibliographystyle{IEEEtran}
\bibliography{gm}

\begin{thebibliography}{10}
\providecommand{\url}[1]{#1}
\csname url@samestyle\endcsname
\providecommand{\newblock}{\relax}
\providecommand{\bibinfo}[2]{#2}
\providecommand{\BIBentrySTDinterwordspacing}{\spaceskip=0pt\relax}
\providecommand{\BIBentryALTinterwordstretchfactor}{4}
\providecommand{\BIBentryALTinterwordspacing}{\spaceskip=\fontdimen2\font plus
\BIBentryALTinterwordstretchfactor\fontdimen3\font minus
  \fontdimen4\font\relax}
\providecommand{\BIBforeignlanguage}[2]{{%
\expandafter\ifx\csname l@#1\endcsname\relax
\typeout{** WARNING: IEEEtran.bst: No hyphenation pattern has been}%
\typeout{** loaded for the language `#1'. Using the pattern for}%
\typeout{** the default language instead.}%
\else
\language=\csname l@#1\endcsname
\fi
#2}}
\providecommand{\BIBdecl}{\relax}
\BIBdecl

\bibitem{file2015res}
{U.S. Energy Information Administration}, ``Residential energy consumption
  survey ({RECS}),'' 2015.

\bibitem{file2015commercial}
------, ``Commercial buildings energy consumption survey ({CBECS}),'' 2015.

\bibitem{jacobs2020unsupervised}
G.~Jacobs and P.~Henneaux, ``Unsupervised learning procedure for nilm
  applications,'' in \emph{2020 IEEE 20th Mediterranean Electrotechnical
  Conference (MELECON)}, pp. 559--564.

\bibitem{angelis2022nilm}
G.-F. Angelis, C.~Timplalexis, S.~Krinidis, D.~Ioannidis, and D.~Tzovaras,
  ``Nilm applications: Literature review of learning approaches, recent
  developments and challenges,'' \emph{Energy and Buildings}, p. 111951, 2022.

\bibitem{cho2018non}
J.~Cho, Z.~Hu, and M.~Sartipi, ``Non-intrusive a/c load disaggregation using
  deep learning,'' in \emph{2018 IEEE/PES Transmission and Distribution
  Conference and Exposition (T\&D)}.\hskip 1em plus 0.5em minus 0.4em\relax
  IEEE, 2018, pp. 1--5.

\bibitem{schirmer2022non}
P.~A. Schirmer and I.~Mporas, ``Non-intrusive load monitoring: A review,''
  \emph{IEEE Transactions on Smart Grid}, 2022.

\bibitem{d2019transfer}
M.~D’Incecco, S.~Squartini, and M.~Zhong, ``Transfer learning for
  non-intrusive load monitoring,'' \emph{IEEE Transactions on Smart Grid},
  vol.~11, no.~2, pp. 1419--1429, 2019.

\bibitem{liang2019hvac}
M.~Liang, Y.~Meng, N.~Lu, D.~Lubkeman, and A.~Kling, ``{HVAC} load
  disaggregation using low-resolution smart meter data,'' in \emph{2019 IEEE
  Power \& Energy Society Innovative Smart Grid Technologies Conference
  (ISGT)}.\hskip 1em plus 0.5em minus 0.4em\relax IEEE, 2019, pp. 1--5.

\bibitem{zhang2018sequence}
C.~Zhang, M.~Zhong, Z.~Wang, N.~Goddard, and C.~Sutton, ``Sequence-to-point
  learning with neural networks for non-intrusive load monitoring,'' in
  \emph{Proceedings of the AAAI conference on artificial intelligence},
  vol.~32, no.~1, 2018.

\bibitem{pecan}
{Pecan Street Inc.}, ``{Pecan Street Dataport},'' \url{https://www.
  pecanstreet.org/dataport/}, 2022.

\end{thebibliography}

\end{document}